\documentclass[conference]{IEEEtran}

\IEEEoverridecommandlockouts
\usepackage{tikz}
\usepackage[cmex10]{amsmath}
\newtheorem{theorem}{Theorem}
\newcommand{\isdef}{\stackrel{\mathrm{def}}{=}}
\DeclareMathOperator{\perm}{perm}

\begin{document}

\title{The Role Model Estimator Revisited}
\author{Jossy Sayir}
\author{\IEEEauthorblockN{Jossy Sayir}
\IEEEauthorblockA{University of Cambridge, U.K.\\
Email: jossy.sayir@eng.cam.ac.uk}
\thanks{Funded in part by the European Research Council under ERC grant
agreement 259663 and by the FP7 Network of Excellence NEWCOM\#
under grant agreement 318306.}}


%


\maketitle

\begin{abstract}
We re-visit the role model strategy introduced in an earlier paper,
which allows one to train an estimator for degraded observations
by imitating a reference estimator that has access to superior 
observations. We show that, while it is true and surprising that
this strategy yields the optimal Bayesian estimator for the degraded
observations, it in fact reduces to a much simpler form in the 
non-parametric case, which corresponds to a type of Monte Carlo
integration. We then show an example for which only parametric
estimation can be implemented and discuss further applications
for discrete parametric estimation where the role model strategy
does have its uses, although it loses claim to optimality in 
this context.
\end{abstract}


%
\IEEEpeerreviewmaketitle

\section{Preamble}

In 2008, I submitted a short semi-technical paper \cite{sayir2008} to
the IT transactions on the
occasion of James L.~Massey's $75^{th}$ birthday. One aim
of the paper was to please Jim who often expressed his liking
for conceptual papers with simple technical content. 
Although the paper was dropped for reasons that will become
apparent, it achieved its aim of pleasing Jim who repeatedly
commented positively on the paper in the years
before he passed away. I would speculate that Jim also liked
the pun on the ``role model'' metaphor in the paper mirrorring
our relationship as past student to PhD advisor.

The present paper re-visits the ideas presented in \cite{sayir2008} and
brings a fresh perspective on the subject. 
In the following section, we will introduce and discuss
the role model strategy. Section~\ref{sec:mcint}
shows how the solution of the role model convex program reduces to
Monte Carlo integration in the non-parametric case, a much simpler
technique well known in the Bayesian community. This realization
is the reason why the original paper
project \cite{sayir2008} was dropped.
Section~\ref{sec:parcase} discusses the parametric case, where the
role model strategy may be of use after all, and why
its relevance was not immediately obvious because
we operate in the domain of discrete probability mass functions where
parametric estimation is not normally considered. An example 
involving the constraint node operation in a factor graph based
SUDOKU solver is presented where parametric estimation is useful,
and other potential applications are discussed.

\section{Introduction and the Role Model Estimator}

\begin{figure}[!h]
\centering
\begin{tikzpicture}
\draw (0,6) rectangle (2,7.5);
\node [align=center] at (1,6.75) {Discrete \\  Memoryless \\ Source};
\draw [->] (1,6) -- (1,5);
\node [right] at (1.1,5.5) {$X_k$};
\draw (0,3.5) rectangle (2,5);
\node [align=center] at (1,4.25) {Discrete \\ Memoryless \\ Channel 1};
\draw [->] (1,3.5) -- (1,2.5);
\node [left] at (.9,3) {$Y_k$};
\draw (0,1) rectangle (2,2.5);
\node [align=center] at (1,1.75) {Discrete \\ Memoryless \\ Channel 2};
\draw [->] (2,1.75) -- (3,1.75);
\node [above] at (2.5,1.85) {$Z_k$};
\draw (3,1) rectangle (5,2.5);
\node [align=center] at (4,1.75) {Estimator in \\ Training};
\draw [->] (1,3) -- (2.5,3) -- (2.5,4.75) -- (3,4.75);
\draw (3,4) rectangle (5,5.5);
\node [align=center] at (4,4.75) {Role Model \\ Estimator};
\draw [->] (5,4.75) -- (6,4.75);
\draw [->] (5,1.75) -- (6,1.75);
\node [align=left,above] at (6,4.75) {$P_{X|Y_k=y}$};
\node [align=left,above] at (6,1.75) {$Q_{X|Z_k=z}$};
\end{tikzpicture}
\caption{The Role Model Framework}
\label{fig:rolemodel}
\end{figure}
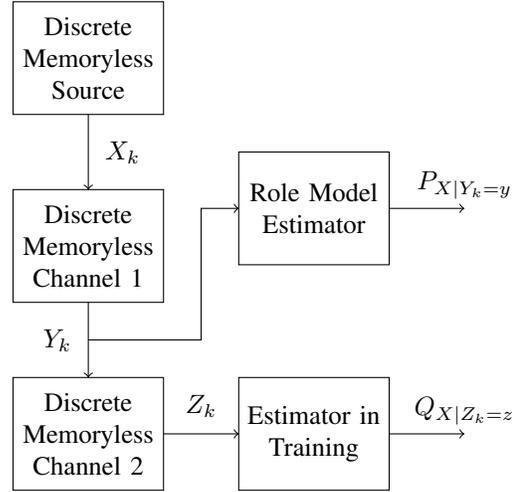

The role model framework introduced in \cite{sayir2008} is illustrated in
Figure~\ref{fig:rolemodel}.
The discrete random variables $X_k$, $Y_k$ and $Z_k$ form a Markov chain
for every $k$. In the following, we drop the time index $k$ when not
essential as our source and channels are assumed to be memoryless. 
The role model estimator is the optimal estimator for $X$ given the 
observation $Y$, which provides for every observation $Y=y$ the full
a-posteriori probability mass function $P_{X|Y=y}$ over the domain of $X$.
Our aim is to train an estimator for $X$ using the random variable $Z=z$,
which is labeled ``estimator in training'' in the figure. The output
of this estimator is labeled $Q_{X|Z=z}$ to reflect the fact that it is
not necessarily the true a-posteriori probability mass function of $X$
given the observation $Z=z$, but an approximation thereof. The estimator
in training is Bayesian optimal if $Q_{X|Z=z}=P_{X|Z=z}$ for every $P_Z(z)>0$.

The reason why $P_{X|Z=z}$ is not available directly may be that
the channel $P_{Z|Y}$ is unknown, or that
the channel $P_{Z|Y}$ is known but that the resulting exact computation
of $P_{X|Z=z}$ is too complex for practical use. The particularity
of the role model framework, in contrast to more complicated estimation
frameworks such as those where the EM and similar algorithms operate,
is that we {\em have access} to the role model estimator
and to its output to help design the estimator in training. This
brings up the justified question of why we don't just use the role
model estimator directly instead of training an estimator based on $Z$.
This may have
several reasons:
\begin{itemize}
\item the observations $Y_k$ and the resulting a-posteriori distributions
  $P_{X|Y=y}$ may only be available during a training phase but not
  when our estimator goes live;
\item the observations $Y_k$ may only be available intermittently and our
  estimator in training is required to fill the gaps at times $k$ when
  $Y_k$ is not available;
\item the computation of $P_{X|Y=y}$ may be too costly and only feasible
  offline during a simulation, or online intermittently for the purpose
  of training the estimator $Q_{X|Z=z}$.
\end{itemize}
We will later discuss a few technical examples in the context of iterative decoding
and communication receivers where these conditions are fulfilled.
\cite{sayir2008} gives hypothetical general examples outside
the domain of communications where this scenario could also be of interest.

What we call the {\em role model strategy} consists in aiming to minimize the expected divergence
between the a-posteriori distribution $P_{X|Y=y}$ computed by the role model estimator,
and the distribution-valued heuristic output $Q_{X|Z=z}$ of the estimator in training,
i.e., to seek the $Q_{X|Z=z}$ for every $z$ that minimizes
\[
ED(P_{X|Y}||Q_{X|Z}) \isdef  \sum_z\sum_y P(yz) D(P_{X|Y=y}||Q_{X|Z=z}),
\]
where we use the notation $ED(.||.)$ as in \cite{coverthomas} to signify the
expected information divergence, where expectation is always taken on the joint
distribution of the conditioning variables.
The averaging required to compute this expression may be impractical, and hence
we use the law of large numbers and the fact that all our processes are ergodic
to state
\begin{equation}
ED(P_{X|Y}||Q_{X|Z}) =  \lim_{N\rightarrow\infty}\frac{1}{N} \sum_{k=1}^N D(P_{X|Y_k=y_k}||Q_{X|Z_k=z_k}),
\label{eq:time-average}
\end{equation}
and approximate the quantity to be minimized by a time average of the
divergence between the two distribution-valued outputs of our estimators. Note
that this may look like a frequentist/empirical approach, but we are at no point counting
frequencies here, so the divergences being averaged are true divergences. It is 
only the average divergence that becomes an approximation if we perform the
time averaging over a finite time interval of length $N$ rather than taking the
limit as $N$ goes to infinity. We note that
$ED(P_{X|Y}||Q_{X|Z})$ is convex in $Q_{X|Z}$, and hence the set of distributions
$Q_{X|Z=z}$ for every $z$ that we need can be sought using numerical convex
optimization techniques.

We devised the role model strategy as a heuristic approach to address this
type of scenario.
We had no expectation that this strategy could be optimal.
The divergence that is minimized cannot in general be reduced to zero,
unless $Z$ is a sufficient statistic for $Y$ with respect to $X$, which is
never the case in the applications of interest. Hence, this is not a system
identification problem, where the estimator in training eventually models
the role model. It therefore came as a surprise when we realized that
the following holds:
\begin{theorem}[The ``role model'' theorem] If $X$, $Y$ and $Z$ form a 
Markov chain $X-Y-Z$, then
\[
ED(P_{X|Y}||Q_{X|Z}) = H(X|Z) - H(X|Y) + ED(P_{X|Z}||Q_{X|Z}).
\]
In particular, 
\[
ED(P_{X|Y}||Q_{X|Z}) \geq H(X|Z) - H(X|Y)
\]
with equality if and only if $Q_{X|Z=z} = P_{X|Z=z}$
for all $z$ such that $P(z)>0$.
\label{th:rolemodel}
\end{theorem}
The theorem shows that the minimization we suggested converges to the
optimal solution $Q_{X|Z}=P_{X|Y}$. Hence, by imitating the 
role model, we converge to the best solution given our degraded observations,
despite the fact that the role model we seek to imitate has better
observations. The proof of the theorem is trivial and given in~\cite{sayir2008}.
Note that the theorem requires the Markov property. A similar-looking result
can be shown when the Markov property does not hold by stating the
identity
\begin{eqnarray*}
\sum_{xyz}P(yz) P(x|yz) \log\frac{P(x|y)}{Q(x|z)} &=& ED(P_{X|Z}||Q_{X|Z}) \\ &&+ H(X|Z) - H(X|Y),
\end{eqnarray*}
effectively showing that $Q_{X|Z}=P_{X|Z}$ minimizes the expression on the left, but this
expression is only equal to $ED(P_{X|Y}||Q_{X|Z})$ when the Markov condition
is verified.

It should be stressed that the appellation ``theorem'' was chosen for this 
result not on the basis of its mathematical intricacy, which it clearly lacks,
but on the basis of its conceptual counter-intuitiveness (from the author's perspective)
and central role it was thought to have in the applications
under consideration. 
In the following section, we will show that the role
model strategy reduces to a much simpler form that is well known in the
Bayesian estimation community, after discussing a class of applications and their
constraints. The role model strategy will regain some meaning in the last section
of the paper, where we show a class of applications where the simpler method does not
apply but where the role model strategy remains a valid approach.

\section{The Non-Parametric Case  and Monte Carlo Integration}
\label{sec:mcint}

\begin{figure}[h!]
\centering
\begin{tikzpicture}
\draw (1,1) rectangle (1.5,1.5) ;
\draw [->] (0.2,0.2) -- (1,1);
\draw [->] (0,1.25) -- (1,1.25);
\draw [->] (0.2,2.3) -- (1,1.5);
\draw [->] (1.5,1.25) -- (2.5,1.25);
\node [align=right,below] at (.9,.7) {$Y_1$};
\node [align=left,below] at (.2,1.25) {$Y_2$};
\node [align=right,below] at (.3,2) {$Y_3$};
\node [align=right,above] at (2.5,1.25) {$P_{X|\underline{Y}}$};
\node [align=right,below] at (2.5,1.2) {$Q_{X|Z}$};
\end{tikzpicture}
\caption{The min-sum approximation for LDPC decoders}
\label{fig:min-sum}
\end{figure}
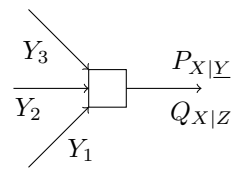

Initial interest for the scenario described was born out of efforts to design
optimal post-processing procedures for sub-optimal components in iterative
decoders. In the min-sum approximation of the sum product algorithm for decoding
low-density parity-check (LDPC) codes, the optimal Bayesian operation under
independence assumption in the
constraint nodes of the decoder is replaced by a sub-optimal operation as
illustrated in Figure~\ref{fig:min-sum}. 
In the figure, the incoming observations $Y_1,Y_2,Y_3$ are aggregated
from channel observations during previous decoder iterations and are
assumed independent, as is common practice in the design of belief
propagation algorithms. 
In this case, the role model estimator, expressed as a mapping of log-likelihood ratios, 
is given by
\[
L(X|\underline{Y}) = 2\tanh^{-1}\left(\prod_i \tanh \frac{L(X|Y_i)}{2}\right),
\]
which is a fairly complex scalar function of multiple variables often considered
too costly for implementation,
while the estimator in training is a function $Q_{X|Z=z}$ of
\[
Z = \left(\min_i|L(Y_i)|,\prod_i\mbox{sign}L(Y_i)\right).
\]
For this simple binary case, the optimal post-processing $P_{X|Z=z}$ can 
be computed analytically \cite{lechner2004} under Gaussian assumption and
is fairly simple to compute when the variances of all incoming observations
are identical. However, as soon as we deviate from this case, i.e., when
the incoming observations have different variances as is the case for irregular LDPC
codes, or if we wish to go beyond the Gaussian simplifying assumption, 
$P_{X|Z=z}$ becomes very difficult to compute. Hence the role model approach
allows us to use numerical optimization algorithms to train a post-processing
function to converge to the optimal estimator by running the 
sum-product rule and the estimator-in-training in parallel offline during a simulation,
and then using the resulting low-complexity estimator online in the device
(e.g., a mobile handset).
Another potential practical scenario consists in running both estimators in parallel
in the device for a limited time while training the low complexity estimator,
then shutting off the more complex estimator to save energy and conserve
battery time. Note that the random variable $Z$ in this example
is continuous but scalar. A fairly accurate estimator can be trained
by quantizing $Z$ finely and computing a lookup table of the a-posteriori
distributions of $X$ for each quantized value of $Z$. 

We will now show that, for this non-parametric approach that aims to estimate
the a-posteriori distributions of $X$ for all values of $Z$, the role model
strategy reduces to a much simpler method well known in the Bayesian
community as a case of Monte Carlo integration.

For now, let us approach the optimization problem via the time-averaging
formulation (\ref{eq:time-average}) where we operate on a finite block
length and drop the limit for simplicity. It is easy to see that the 
minimization with respect to the matrix $Q_{X|Z}(x|z)$ for all $x$ and $z$ that we
require simplifies to separate maximizations for each individual $z$ of the
type
\[
\begin{cases}
\max_{Q(.|z)} &\sum_{k:z_k=z}\sum_x P(x|y_k)\log Q(x|z) \\
\text{subj.~to} &\sum_xQ(x|z) = 1 \\
&Q(x|z)\geq 0, \forall x
\end{cases}
\]
We now take the liberty of ignoring the inequality constraints and
setting up the Lagrange conditions rather than the KKT conditions,
because the solution will show that there is no danger of any 
variables becoming negative.
For any $z$, we obtain by differentiating with respect to $Q(x|z)$
\[
\sum_{k:z_k=z} \frac{P(x|y_k)}{Q(x|z)} = \lambda
\]
and hence
\[
Q(x|z) = \lambda^{-1} \sum_{k:z_k=z} P(x|y_k).
\]
The normalization condition requires that
$\lambda = \vert \{k:z_k=z\}\vert$ and the solutions
clearly satisfy $Q(x|z)\geq 0$ since they are obtained
as an average of probabilities.

We conclude that the solution of the role model strategy for any $z$ in the time
averaging case is simply the time average of the a-posteriori distributions
computed by the role model for all $Y_k$ such that $Z_k=z$. Again, we insist
that this is not simply a frequentist/empirical approach as may appear. The quantities
being added here are not numbers of occurrences but true Bayesian a-posteriori
distributions computed by the role model. The correct training for our estimator
of $X$ for the symbol $z$ is to average the distribution-valued estimations
of the role model componentwise over the time instances when $z$ is observed. 

Although we showed this for finite $N$, it is easy to see that the same
holds in the limit as $N$ goes to infinity, and hence for the expectation
$ED(P_{X|Y}||Q_{X|Z})$. An alternative view is that the optimal strategy 
is to evaluate the sum
\[
P(x|z)=\sum_yP(x|y)P(y|z)=E_{P_{Y|Z=z}}[P(X|Y)]
\]
as a time average, as briefly stated in \cite{sayir2010-isit}.

In the min-sum algorithm discussed above, and in any similar
applications where it is possible to adapt the full $\vert\mathcal{Z}\vert\times\vert\mathcal{X}\vert$
parameter set for $Q_{X|Z}$, the role model strategy is an overkill
and Monte Carlo integration gives the same solution by elementary
averaging without resorting to complicated numerical convex
optimization methods. In the next section, we will see that
there is still a niche for the role model strategy when the
full parameter set is too large for practice.

\section{The Parametric Case: an Example}
\label{sec:parcase}

When the full parameter set is not available, Monte Carlo integration
is not an option and the role model strategy becomes a possibly interesting
approach. While this is easy to state, it is not an obvious proposition 
because we don't tend to think of parametric estimation for discrete
random variables. Indeed, we are not proposing to constrain the conditional
probability mass functions $Q_{X|Z}$ to be parametric distributions in the
sense that a Gaussian density is a parametric probability density function.
Rather, as we will see in our examples, there are scenarios where the 
domain $\mathcal{Z}$ of $Z$ makes it impractical to estimate an a-posteriori
model $Q_{X|Z=z}$ for every possible $z$. In such scenarios, we may be 
constrained to using a parametric function of $Z$, i.e., $Q_{X|Z=z}=f_\alpha(z)$. 
In such a case, the role model strategy loses its optimality as the 
space of possible functions $f_\alpha(.)$ will not in general include the
mapping that makes $Q_{X|Z}$ converge to $P_{X|Z}$. Hence, the role
model strategy in this context is a purely heuristic approach that may
or may not exhibit advantages or weaknesses with respect to other
heuristic optimization
criteria and can be judged solely on the basis of its numerical performance.

An early example applying the role model strategy in a semi-parametric manner
was described in \cite{sayir2010-turbo} for a hypothetical rank-based 
message-passing decoder for non-binary LDPC codes. In fact, a more pertinent
question than that studied in \cite{sayir2010-turbo} would be to design post-processing
operations for the suboptimal operations in the Extended Min-Sum (EMS) algorithm \cite{declercq2006},
a reduced complexity version of the sum-product algorithm for non-binary LDPC
codes. However, the EMS algorithm is quite a difficult construct to understand, so
that a full study of parametric post-processing, while practically relevant,
would obscure rather than clarify matters in the context of this paper. Hence,
we have chosen to treat an alternative example of lesser practical relevance
but that is easier to understand.

\begin{figure}[!h]
\centering
\includegraphics[width=\columnwidth]{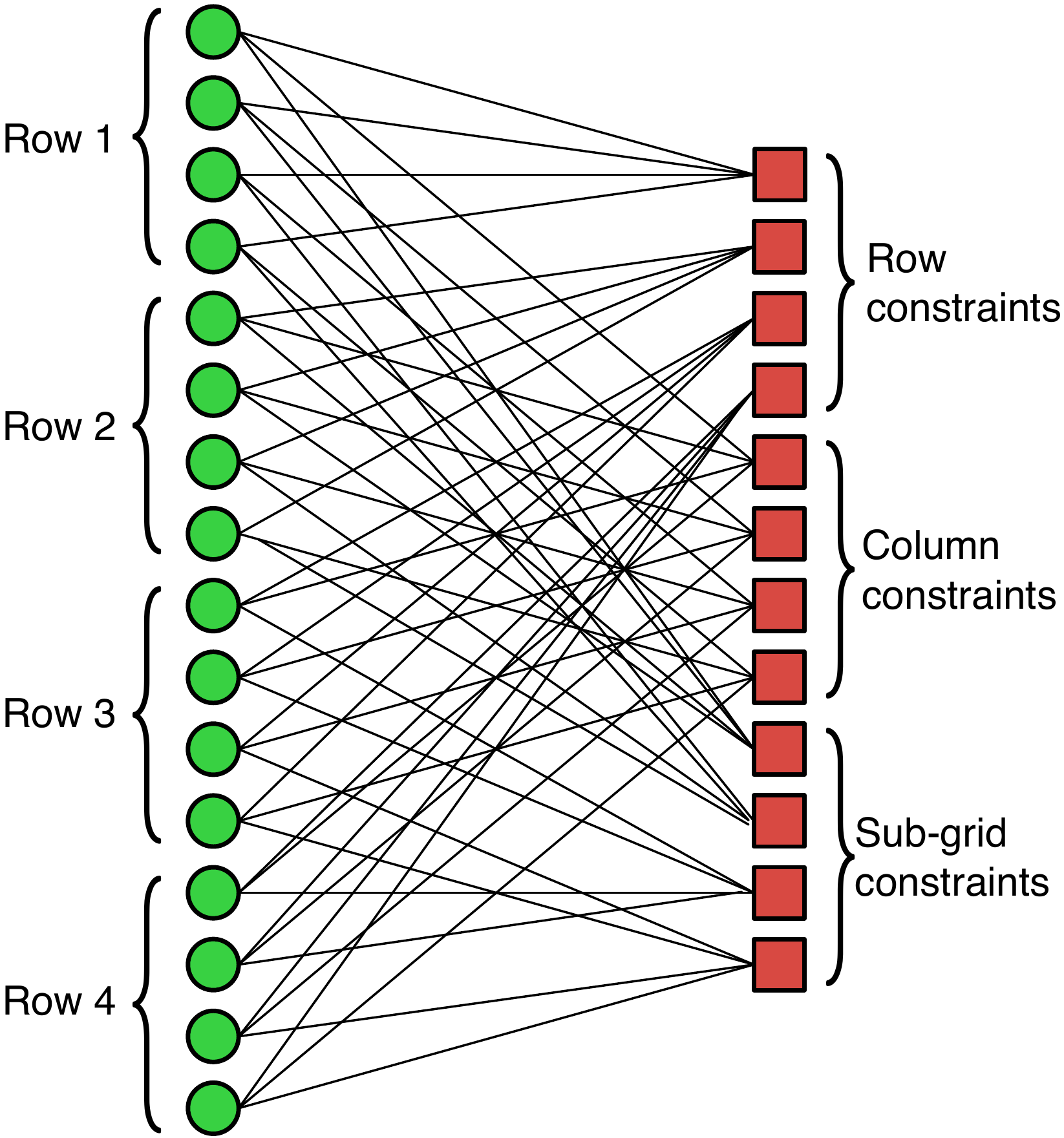}
\caption{The factor graph of a $4\times 4$ SUDOKU solver}
\label{fig:sudoku}
\end{figure}

The example is the use of graph-based decoding for solving soft SUDOKU puzzles.
We omit an introduction to universally known SUDOKU puzzles and refer the 
reader to \cite{sudoku} for futher details and definitions.
By ``soft'' SUDOKU, we mean puzzles that receive
general noisy observations of the correct entries in the grid rather than
observations that are either correct or erased. Observations for every 
entry in the grid are available as a-posteriori probability
mass functions over the 9-ary alphabet using a known and accurate channel model.
SUDOKU puzzles can be represented as a factor graph where every one of the 81
variables is connected to 3 constraints and every constraint (9 rows, 9 columns
and 9 subgrids) involves 3 variables. The factor graph of a $4\times 4$ SUDOKU
defined over the alphabet $\{1,2,3,4\}$ is represented in Figure~\ref{fig:sudoku}.
Our interest in the context of this paper
is for the operation in the constraint nodes. Constraint nodes receive 9 
a-posteriori observations of their participating variables, which we assume
to be independent in line with common practice in belief propagation algorithms.
The constraint node's task is to return to each variable its a-posteriori 
probability given the observations of the remaining 8 variables in the
constraint. Let us denote by $\mathbf{M} = [m_{ij}]$ the $9\times 9$ matrix
of incoming messages into a constraint node, where 
\[
m_{ij} = P(X_i=j|\underline{Y}_i)
\]
where $\underline{Y}_i$ generically denotes the set of channel observations
that led to the incoming message on the $i$-th branch into the constraint
node. It is clear that the probability that the first variable in the constraint
has value $j$, given observations of the other 8 variables, is the sum of
probabilities of all configurations of the other 8 variables that don't 
include the value $j$. If we denote by $\mathbf{M}_{\setminus ij}$ the matrix
$\mathbf{M}$ with its $i$-th row and $j$-th column removed, we can 
state that an outgoing message component from the constraint node can 
be expressed as
\[
m'_{ij} = \perm(\mathbf{M}_{\setminus ij})
\]
where $\perm(\mathbf{A})$ denotes the Cauchy permanent \cite{permanent}
of the matrix $\mathbf{A}$. The permanent is a patently difficult 
function to compute and the best algorithms known compute an approximation
of the permanent in probabilistically polynomial time, which polynomial
is considerably larger than $n!$ for sizes $n$ of interest to us. We can 
hence assume that $8!=40320$ operations are needed to compute the permanent
above. This is a large number of operations for every node at every
iteration of a belief propagation solver, but well within the range of 
offline simulation, so a perfect testing ground for our role model strategy.
We can now try to replace the permanent computation by any approximation
and use the role model strategy to design post-processing functions for the
approximation.

\begin{figure}[!h]
\centering
\includegraphics[clip=true,trim=280 50 260 40,width=\columnwidth]{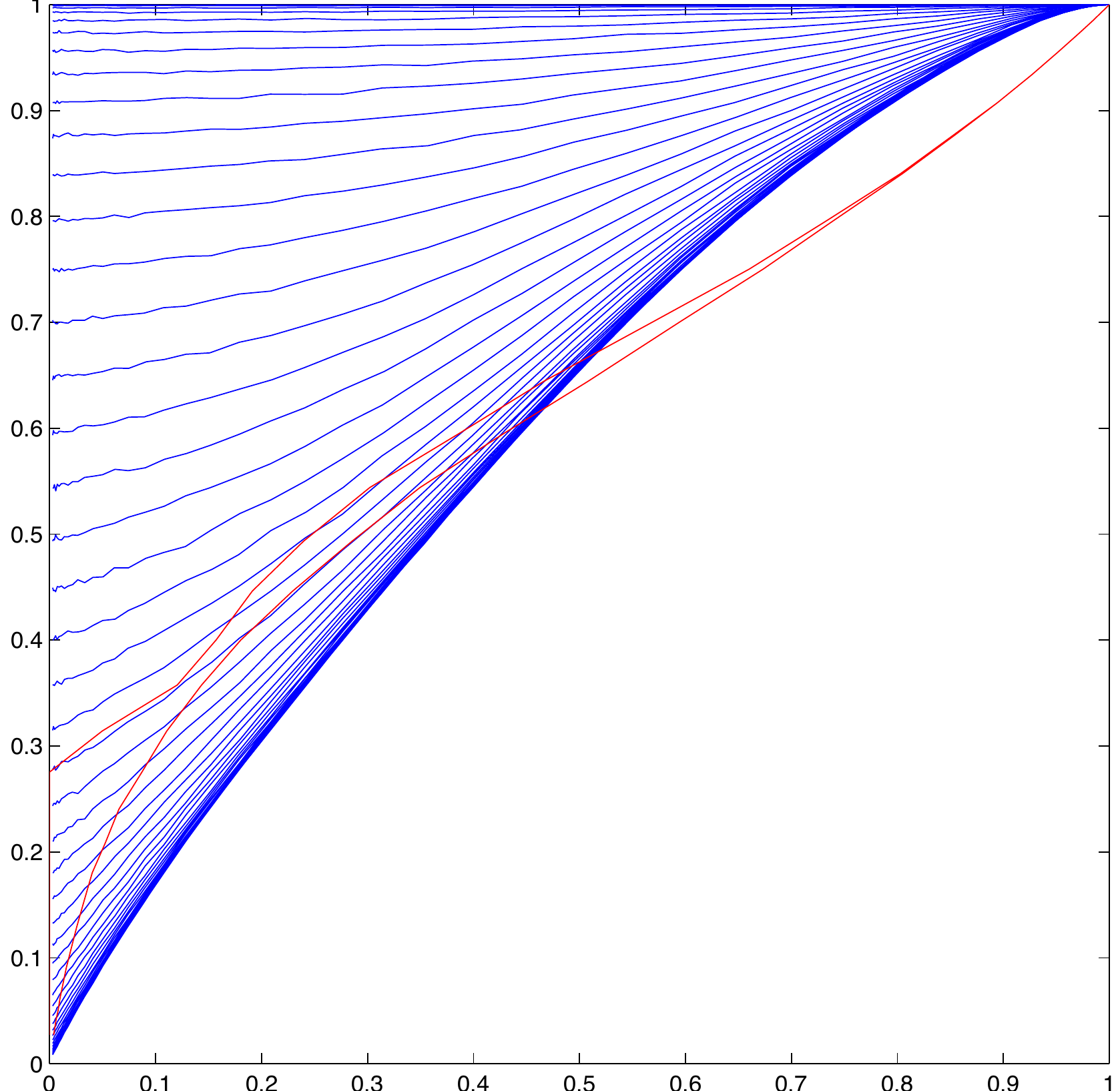}
\caption{EXIT chart of a SUDOKU solver using the optimal and approximate constraint node operations}
\label{fig:EXIT}
\end{figure}

For example, we can opt to do the following:
\begin{itemize}
\item take the 3 largest elements in each row of $\mathbf{M}$ and replace
  the remaining entries by a uniform distribution adding to the same sum
  to obtain the matrix $\mathbf{M'}$,
\item re-write the matrix $\mathbf{M'}$ as $\mathbf{H}+\mathbf{T}$ where
  $\mathbf{T}$ contains uniform rows whose values are consistent with the uniform
  tails produced in the previous step, and $\mathbf{H}$ contains the 
  non-uniform values minus the uniform tail value for those head entries not
  in the tails, and zero where the tail entries of $\mathbf{M'}$ are;
\item we now approximate the required permanent as 
  \[
\perm\mathbf{M}_{\setminus ij} \approx \perm\mathbf{H}_{\setminus ij} + \perm\mathbf{T}_{\setminus ij}
\]
\item we compute the elements of the outgoing matrix using this approximation
and normalize the rows so they sum to 1 and look like true probability mass
functions.
\end{itemize}
This is much easier to compute because $\mathbf{H}$ is sparse and $\mathbf{T}$ 
has uniform rows, but it is a very poor approximation because the permanent of 
a sum of matrices is not at all well approximated by the sum of their permanents.
Figure~\ref{fig:EXIT} shows the EXIT chart of a factor-graph
based SUDOKU solver, where the two red curves correspond to the optimal and
sub-optimal constraint node operations, and the blue curves correspond to the
constraint node operations for various channel signal to noise ratios (SNR). 
Surprisingly, the red curves are not too far apart in particular in the top
half of the EXIT chart, indicating that despite the very rough approximation
we are using, the result is sufficiently informative to achieve acceptable
performance, and the full complexity permanent computation should only be used
in the early iterations. 

Now for the application of the role model strategy. The observations for our
role model postprocessor in this case consist of the rows of the outgoing
matrix computed using the permanent approximations. The observation
space is the set of 9-ary probability distributions. This is a continuous space and
is no longer scalar like in the binary min-sum case. Hence, we cannot simply
quantize it finely in order to apply Monte Carlo integration and converge to the
optimal a-posteriori estimation. What we can do, however, is to apply arbitrary
transformations to the probability vectors. For example, we can take 
replace the sum $\perm\mathbf{H}+\perm\mathbf{T}$ by a weighted sum 
 $\alpha_i\perm\mathbf{H}+(1-\alpha_i)\perm\mathbf{T}$ and
optimize the weights $\alpha_i$.
Hence the problem becomes 
one of finding the best parameters $\alpha_i$ to optimize the solver performance.
The problem is that solver performance itself is difficult to measure and
can only be optimized by exhaustive search algorithm. The role model
strategy in this case yields a tracktable convex optimization procedure
where $ED(P_{X|Y}||Q_{X|Z})$ is the optimization metric. $P_{X|Y}$ here 
is the correct a-posteriori distribution obtained with the true 
permanent, and $Q_{X|Z}$ is the $\alpha_i$-corrected result of the
sum of permanents approximation. Note that with this approach, we have lost
any claim of optimality, and anyone who prefers another metric over ours
is entitled to do so. The only valid criterion for comparing metrics is simulated
performance of the resulting optimized solvers.

\section{Conclusion}

We have described the role model strategy as a convex program whose solution
is the Bayesian optimal estimator in training. We
showed that the strategy reduces to Monte Carlo integration in the non-parametric
case, and discussed the parametric case with an example where the strategy can
be used but Monte Carlo integration would not work.

In fact, applications of post-processing optimization for 
sub-optimal estimators are burgeoning in the literature and many
metrics have been proposed for optimizing the post-processing stage of,
say, the EMS algorithm for non-binary LDPC codes, demodulators for 
Bit-Interleaved Coded Modulation (BICM) and many others. Some, such as
\cite{nguyen2011} claim theoretical motives for
their approaches, while others, such as \cite{szczecinski2012},
are self-declaredly heuristic in their approach. Given our analysis
so far and the fact that these are all parametric models, we tend
to agree with the latter.

\section*{Acknowledgment}

I learned that ``my'' role model strategy reduces to elementary Monte Carlo integration
upon joining the University of Cambridge in a conversation with my then officemate
and now friend Simon Hill, a fact that earns him my warmest gratitude as well as my
sincere apologies for the inappropriate language he may have heard
when I found out. I also wish to thank to the then associate editor of the IT 
transactions Michael Gastpar who handled my submission and the three gracious anonymous reviewers
who I realize put considerable effort into providing constructive feedback,
with apologies for never writing back to explain that I was not revising the paper as a result
of the conversation mentioned above.




\begin{thebibliography}{10}
\providecommand{\url}[1]{#1}
\csname url@samestyle\endcsname
\providecommand{\newblock}{\relax}
\providecommand{\bibinfo}[2]{#2}
\providecommand{\BIBentrySTDinterwordspacing}{\spaceskip=0pt\relax}
\providecommand{\BIBentryALTinterwordstretchfactor}{4}
\providecommand{\BIBentryALTinterwordspacing}{\spaceskip=\fontdimen2\font plus
\BIBentryALTinterwordstretchfactor\fontdimen3\font minus
  \fontdimen4\font\relax}
\providecommand{\BIBforeignlanguage}[2]{{%
\expandafter\ifx\csname l@#1\endcsname\relax
\typeout{** WARNING: IEEEtran.bst: No hyphenation pattern has been}%
\typeout{** loaded for the language `#1'. Using the pattern for}%
\typeout{** the default language instead.}%
\else
\language=\csname l@#1\endcsname
\fi
#2}}
\providecommand{\BIBdecl}{\relax}
\BIBdecl

\bibitem{sayir2008}
\BIBentryALTinterwordspacing
J.~Sayir, ``What makes a good role model,'' 2008, arXiv pre-print. [Online].
  Available: \url{http://arxiv.org/abs/0809.1300}
\BIBentrySTDinterwordspacing

\bibitem{coverthomas}
T.~M. Cover and J.~A. Thomas, \emph{Elements of Information Theory}.\hskip 1em
  plus 0.5em minus 0.4em\relax Wiley, New York, 1991, no. ISBN 0-471-24195-4.

\bibitem{lechner2004}
G.~Lechner and J.~Sayir, ``Improved sum-min decoding of {LDPC} codes,'' in
  \emph{Proc. Int. Symp. Inf. Theory and Its App. ({ISITA})}, Parma, Italy,
  Oct. 2004.

\bibitem{sayir2010-isit}
\BIBentryALTinterwordspacing
J.~Sayir, ``{EXIT} chart approximations using the role model approach,'' in
  \emph{Proc. {IEEE} Int. Symp. Inform. Theory ({ISIT})}, Jun. 2010. [Online].
  Available: \url{http://arxiv.org/abs/1006.0659}
\BIBentrySTDinterwordspacing

\bibitem{sayir2010-turbo}
------, ``Design of non-binary decoders using the role model framework,'' in
  \emph{Proc. Int. Symp. on Turbo Codes \& Rel. Topics}, Brest, France, Sep.
  2010.

\bibitem{declercq2006}
\BIBentryALTinterwordspacing
D.~Declercq and M.~P. Fossorier, ``Decoding algorithms for nonbinary {LDPC}
  codes over {GF(q)},'' \emph{{IEEE} Trans. Commun.}, vol.~55, no.~4, pp.
  633--643, Apr. 2007. [Online]. Available:
  \url{http://publi-etis.ensea.fr/2007/DF07"}
\BIBentrySTDinterwordspacing

\bibitem{sudoku}
\BIBentryALTinterwordspacing
``Sudoku,'' article in Wikipedia. [Online]. Available:
  \url{http://en.wikipedia.org/wiki/Sudoku}
\BIBentrySTDinterwordspacing

\bibitem{permanent}
\BIBentryALTinterwordspacing
``Permanent,'' article in Wikipedia. [Online]. Available:
  \url{http://en.wikipedia.org/wiki/Permanent}
\BIBentrySTDinterwordspacing

\bibitem{nguyen2011}
T.~T. Nguyen and L.~Lampe, ``Bit-interleaved coded modulation with mismatched
  decoding metrics,'' \emph{{IEEE} Trans. Commun.}, vol.~59, no.~2, pp.
  437--447, feb 2011.

\bibitem{szczecinski2012}
L.~Szczecinski, ``Correction of mismatched l-values in bicm receivers,''
  \emph{{IEEE} Trans. Commun.}, vol.~60, no.~11, pp. 3198--3208, nov 2012.

\end{thebibliography}
%



\end{document}